# High critical current density in textured Ba-122/Ag tapes fabricated by scalable rolling process

Chiheng Dong[a], Chao Yao[a], He Lin[a], Xianping Zhang[a], Qianjun Zhang[a], Dongliang Wang[a], Yanwei Ma,[a,*] Hidetoshi Oguro[b], Satoshi Awaji[b], Kazuo Watanabe[b]

[a]*Key Laboratory of Applied Superconductivity, Institute of Electrical Engineering, Chinese Academy of Sciences, Beijing 100190*
[b]*High Field Laboratory for Superconducting Materials, Institute for Materials Research, Tohoku University, Sendai 980-8577, Japan*

The industrial manufacturing of the long-length iron-pnictide wires and tapes requires simple and low-cost technology. Although the transport critical current density of the FeAs-122 tapes has already achieved the practical application level, their fabrication procedures are relatively complicated or required specialized instruments. In this paper, we fabricated the Ag-sheathed $Ba_{0.6}K_{0.4}Fe_2As_2$ tapes by scalable rolling process. The critical current density $J_c$ at 4.2 K and 10 T has achieved $5.4 \times 10^4$ A/cm$^2$. We ascribe the excellent performance to the high purity and homogeneity of the superconducting phase, strong flux-pinning, high density and $c$-axis texture of the superconducting core. Our method provides a simple way to scale up the production of the long-length iron-based tapes with high $J_c$.

[*]Electronic mail: ywma@mail.iee.ac.cn

The discovery of iron pnictide superconductors[1] is a triumph in the superconducting science. Among the varieties of newly found iron-based superconductors, FeAs-122 superconductor intrigues great interest because of its relatively high superconducting transition temperature[2], high upper critical field[3,4] $H_{c2}$ as well as the small anisotropy[5,6], which imply its potential application in the superconducting power cables and high field magnets. However, initial enthusiasm devoted to the polycrystalline samples was not so successful because of the existence of electromagnetic granularity[7,8] and no bulk scale supercurrent was observed. Subsequent FeAs-1111 polycrystals synthesized by the high pressure method exhibited global current in the bulk sample, though its magnitude was much lower than the intragranular current[8]. Further studies on the epitaxial films indicated that the misalignment of the crystalline orientation greatly suppresses the critical current density[9,10] with a critical angle $\theta_c \sim 9^\circ$. On the other hand, the grain boundary wetting phase FeAs[11], the amorphous layers as well as the oxide layers[12] between the grains strongly suppress the superconducting order parameter at the grain boundary. The cracks[11], pores and voids observed in the bulk samples decrease the actual area of the current and become a fatal factor to the supercurrent density.

To overcome these obstacles, powder in tube (PIT) method was applied to fabricate the iron-based superconducting tapes[13]. The cold deformation process was repeatedly optimized to obtain a denser superconducting core with higher $c$-axis texture, the metal additives were introduced to improve the grain connectivity[14,15], hot isostatic pressing and low temperature sintering methods were also adopted to obtain high purity and nearly 100% dense superconducting cores[16,17]. Recently, cycles of rolling and intermediate annealing method was applied to enhance the critical current density of the $(Ba,K)Fe_2As_2$ tapes. Subsequent uniaxial pressing further increased $J_c$ to a higher level. Densification and change of crack structure were attributed to this enhancement. Despite the cold pressing process (CP)[18,19], hot pressing (HP) was also found highly efficient to reduce the number of the pores[20] and increase the core density. Very recently, Zhang *et al.* fabricated the $Sr_{1-x}K_xFe_2As_2$ superconducting tapes via the HP method and improved the transport $J_c$ up to 0.1 MA/cm$^2$ at 10 T and 4.2 K, surpassing the threshold for the practical application[21]. However, as for the CP and HP method, high pressure ranged from 10 Mpa to 4 Gpa[18,20-22] is applied during the manufacture process. Specialized equipments are indispensable to the long length wire production. High-priced production cost is another disadvantage to be concerned. Therefore, an easy and simple process is still required to balance the high performance of the superconducting tapes and the production cost. In this paper, we firstly synthesized the precursor with a new two steps method and then fabricated the Ag-sheathed $Ba_{0.6}K_{0.4}Fe_2As_2$ tapes with one cycle of swaging, drawing and rolling. Transport critical current density $J_c \sim 5.4 \times 10^4$ A/cm$^2$ at 10 T and 4.2 K is achieved.

The quality of the precursor are of great significance to the final current carrying ability of the FeAs-122 superconducting tapes[16,20,23]. However, the purity of the $A_{1-x}K_xFe_2As_2$ (A=Sr, Ba)polycrystal as well as the homogeneity of Sr/Ba and K elements are particularly delicate to the oxygen contamination, high vapor pressure and the loss of the volatile elements. Different from the previous synthesis therapy which takes the elementary substances Sr/Ba and K as the starting material, we used the intermediates BaAs and KAs in order to avoid the above mentioned disadvantages. The high energy ball milling was also applied to throughly mix the materials and obtain the fine particles with sub-micron size[17]. High purity K pieces (99.95 %) and As powders (99.95 %) with ratio 1:1 were mixed together, sealed in a silicon tube and finally sintered at a moderate temperature. Similar process was applied to Ba rod (99.9 %) and As powders. These intermediates were mixed with Fe (99.99 %) and As powders according to the nominal composition: $Ba_{0.6}K_{0.5}Fe_2As_2$. Excess KAs was added to compensate the K loss during the precursor sintering and following tape annealing process. The mixture was throughly ground with a ball milling machine, sealed in a silicon tube and finally heated to 900 ℃ for 35 hours. The as-sintered bulk was ground into powders, uniformly mixed with 5 wt. % Sn powders and packed into a Ag tube with OD 8 mm and ID 5 mm. The Ag tube went through one cycle of swaging, drawing and flat rolling process into the tapes with 0.3 mm thick. The tapes were finally annealed at 900 ℃ for half an hour.

The X-ray diffraction (XRD) patterns of the rolled tapes and the random powders were performed on a Bruker D8 Advance X-ray diffractometer. Analysis of the diffraction data was performed using the GSAS suit of the Rietveld programmes[24,25]. The cross section and the plane of the rolled tapes were well polished before the characterization using the field emission scanning electron microscope (SEM, Zeiss SIGMA). Energy dispersive X-ray spectroscopy was obtained on an EDAX equipment attached to the SEM. We also measured the Vickers hardness using the Vicker hardness tester (401MVD, Wolpert Wilson Instruments) on the polished cross sections with 0.05 kg load and 10s duration in a row at the center of the superconducting cores. The transport critical current $I_c$ at 4.2 K under a magneticfield range from 0 to 14 T was measured using the four probes method with a criterion of 1 $\mu$V/cm at the High Field Laboratory for Superconducting Materials (HFLSM) at Sendai. The transport current $I_c$ was performed on several tapes to ensure the reproducibility. Resistivity and magnetization of the superconducting core were measured via a Physical Property Measurement System (PPMS-9) using the standard four probes method and the vibrating sample magnetometer (VSM), respectively.

The purity of the precursor powder was characterized by the XRD measurement as shown in Fig.1(a). All the diffraction peaks can be well indexed with the I4/*mmm* space group. No trace of impurity phase is observed, indicating that our precursor is of high purity. Fig.1(b) shows the temperature dependence of magnetization for the precursor powder under a 30 Oe magnetic field with a zero field cooling (ZFC) and a field cooling (FC) procedure. The superconducting transition begin at $T_c^{mag} \sim 38$ K. The SEM image of the ground precursor powders is shown in Fig.1(c). Large laminar BaK-122 crystals with different sizes randomly distributed in the figure. Fig.1(d) is the enlarged view of the platelets. The layered structure of the crystal can be clearly observed. The dimension of a typical crystal is about 6.7×8.8 $\mu m^2$. We also performed the EDS mapping to the bulk precursor (not shown here). The constituent elements homogeneously distribute and the average composition is $Ba_{0.55}K_{0.45}Fe_2As_2$.

The XRD patterns for the random powders obtained by grinding the superconducting cores is shown in Fig. 2 (a). The Rietveld refinement is quite successful as indicated by the fairly good reliability factors concluded in table I. According to the results, the main phase of the superconducting core is determined to be BaK-122 (fraction: 98.4 %). The additional diffraction peaks come from Ag which infiltrated inside the cores from the sheath[26]. The lattice parameters of the main BaK-122 phase are $a=b=3.9086(5)$ Å, $c=13.3088(1)$ Å, which is very close to that of $Ba_{0.6}K_{0.4}Fe_2As_2$[2,27]. The fitted chemical formula also indicates that the composition of the superconducting core is in the optimal K doping region. We also measured the composition via the EDS on at least 10 points of the superconducting core. We found that the average composition is $Ba_{0.58}K_{0.42}Fe_2As_2$ which is close to the fitted formula. The content of K is little less than the precursor due to the loss of the alkali metal element during the final annealing process. The XRD profiles of the core surface obtained by mechanically removing the Ag sheath is presented in Fig. 2 (b). Both the BaK-122 phase and the Ag minority phase are observed in the profile and the peak position is exactly the same as that of the random powders. Moreover, the intensities of the (00*l*) peaks for the FeAs-122 phase are strongly enhanced, indicating that the degree of *c*-axis texture is greatly improved after the cold-deformation.

In order to quantitatively evaluate the degree of texture, we calculated the $c$-axis orientation factor F by the Lotgering method[28]:

$$F = \frac{\rho - \rho_0}{1 - \rho_0} \qquad (1)$$

where

$$\rho = \frac{\sum I(00l)}{\sum I(hkl)}, \rho_0 = \frac{\sum I_0(00l)}{\sum I_0(hkl)} \qquad (2)$$

$I$ and $I_0$ are the intensities of each reflection peak for the oriented and random samples, respectively. To obtain an accurate value of the intensities, we firstly subtract the background of the XRD patterns. The calculated F value for the flat rolled tapes is 0.69, which is much larger than that of the $Sr_{1-x}K_xFe_2As_2$ tapes (0.32-0.35 for the flat rolled Ag tapes in Ref.20, 0.565 for the flat rolled iron tapes in Ref.29, 0.52 for the hot pressed Ag tapes in Ref.21). The large F value for the flat rolled Ag-sheathed $Ba_{0.6}K_{0.4}Fe_2As_2$ tape indicates that the traditional PIT method, which is composed of swaging, drawing and flat rolling procedures, can effectively align the planar Ba-122 crystals in the $c$-axis.

The magneto-resistance measurement result is shown in Fig. 3. The resistivity at 300 K and 39 K is 0.32 mΩ cm and 0.038 mΩ cm respectively. The residual resistance ratio is 8.4, manifesting a good grain connectivity. At zero field, the superconducting transition started at $T_c^{onset}$~38.1 K and the superconducting transition width $\Delta T_c$ at zero field is 0.55 K, indicating good quality of our sample. The $T_c$ decreases with increasing magnetic field, resulting a gradually broaden transition. Inset (d) presents the field dependence of $\Delta T_c$. The transition width linearly increases with increasing field. The slope $d\Delta T_c/dH$ is 0.186 K/T. This relatively robust superconductivity under the magnetic field suggests of a potential magnet application. The H-T phase diagram summarized from the magnetoresistance measurement is shown in Fig. 3(b). The $B_{irr}$-T curve is quite close to the $B_{c2}$-T curve, indicating that the flux lattice melting occurs in a narrow region of the H-T phase diagram.

We also notice a tail structure at low temperature close to $T_c$ when applying a magnetic field, implying a possible thermal activated flux flow (TAFF) behavior. According to the TAFF model, the resistivity is:

$$\rho(T, H) = \frac{2\rho_c U}{T} e^{-\frac{U}{T}} \qquad (3)$$

where U is the thermally activated energy. Assuming that $2\rho_c U/T$ is a temperature independent

constant, noted as $\rho_{0f}$, $U=U_0(1-T/T_c)$, then Eq.(3) can be simplified to the Arrhenius relation:

$$ln\rho(T,H) = ln\rho_{0f} - \frac{U_0(H)}{T} + \frac{U_0(H)}{T_c} = ln\rho_0 - \frac{U_0(H)}{T} \qquad (4)$$

Where

$$ln\rho_0 = ln\rho_{0f} + \frac{U_0(H)}{T_c} \qquad (5)$$

. We plot the electrical resistivity $\rho$ as a function of 1/T on a semilogarithmic scale at the magnetic field range from 0 to 9 T, as shown in Fig. 3(e). It is obvious that the Arrhenius relation holds well in a large temperature range, indicating that the TAFF model fits to the flux motion in our $Ba_{0.6}K_{0.4}Fe_2As_2$ tapes. According to Eq.(4), the thermal activated energy can be evaluated from the slope $U_0(H)=-dln\rho/d(1/T)$. The thermal activated energy at 1T is 6764 K, which is close to that of the hot pressed $Sr_{1-x}K_xFe_2As_2$ tapes[21]. We plot $ln\rho_0$-$U_0$ in Fig. 3(f), and find that it can be fitted linearly according to Eq.(5). The values of $\rho_{0f}$, which is deduced from the intercept with the ordinate, is 24.36 mΩ cm. The reciprocal of the slope is 38.5 K which is close to $T_c^{onset}$.

The susceptibility of the superconducting core obtained by peeling off the Ag sheath is measured with a 30 Oe magnetic field parallel to the tape plane in order to reduce the demagnetization effect. As shown in the inset of Fig. 4, the superconducting transition is rather sharp ($T_c^{mag}$~37.5 K, $\Delta T_c^{mag}$~4K), indicating a bulk scale shielding current. The volume fraction of superconductivity is nearly 100 % except for a small proportion of nonsuperconducting phase which is consistent with the XRD results. The rather flat FC branch manifest that the flux pinning strength is considerably strong in our $Ba_{0.6}K_{0.4}Fe_2As_2$ tapes.

The magnetic field dependence of the transport critical current density $J_c$ for the $Ba_{0.6}K_{0.4}Fe_2As_2$ tapes is shown in Fig. 4. The $I_c$ value in self field and low magnetic field is too large to be measured with a 300 ampere current source. Therefore, we only present the data above 4 T. The area of the cross section is calculated using the optical images, as shown in Fig. 5(a). The critical current density at 4 T approaches $7\times10^4$ A/cm$^2$ and decreases slowly with increasing magnetic field. The $J_c$ at 4.2 K and 10 T achieves $5.4\times10^4$ A/cm$^2$, much larger than that of the $Ba_{1-x}K_xFe_2As_2$ tapes[22] fabricated with cycles of rolling and intermediate heating method. We also measured the Vickers microhardness (Hv) on several points on the cross section of the $Ba_{0.6}K_{0.4}Fe_2As_2$ tapes. The average $H_v$ value is 136, much larger than that of the flat rolled Sr-122 tapes ($H_v$~61.2)[20] and the flat rolled Ba-122 tapes[19]

($H_v$~94.0), as well as the uniaxial pressed Ba122 tapes in Ref.19.

Fig. 5 shows the optical and SEM images of the cross section, longitudinal section as well as the core plane of the $Ba_{0.6}K_{0.4}Fe_2As_2$ tapes. Fig. 5(b) is an optical image of the cross section magnified by 500 times. Some voids and pores randomly distributed across the transverse section. The white particles are the precipitated Ag. The SEM image also exhibits the porous structure as well as some cracks about 10 $\mu$m long as shown in Fig. 5(c). The SEM image of the well polished plane of the core is shown in Fig. 5(d). Although the voids and pores still exist on the tape plane, they are much sparser than that of the flat rolled FeAs-122 tapes[20,22]. (e) and (f) shows the longitudinal section of the superconducting core. The BaK-122 crystal planes align along the tape axis (red arrow) due to their two dimensional nature. The length of two typical crystals marked in (f) are 6.3 $\mu$m and 7.3 $\mu$m which is close to that of the precursor.

To summarize, we synthesized the high quality $Ba_{0.6}K_{0.4}Fe_2As_2$ precursor with a new two steps method and fabricated the Ag sheathed tapes by scalable rolling process. The transport critical current density at 4.2 K and 10 T is as high as $5.4 \times 10^4$ A/cm$^2$. We attribute the excellent performance of our tapes to the high purity and homogeneity of the superconducting phase, strong flux pinning, high density and *c*-axis texture of the superconducting core. Higher critical current density can be optimistically expected if the process parameters of our method are optimized to get a superconducting tape with higher density and texture. Our fabrication method provides a reasonable solution to the scalable production of long-length iron-based superconducting tapes with high critical current density.

This work was partially supported by the National 973 Program (Grant No. 2011CBA00105), the National Natural Science Foundation of China (Grant No. 51025726, 51320105015 and 51402292).

**Captions:**

FIG. 1. (a) XRD pattern for the precursor powders, the crystal indices are also marked. (b) Magnetization as a function of temperature measured in a 30 Oe magnetic field with a zero field cooling (ZFC) and field cooling (FC) process. (c) and (d) are the SEM images for the precursor powders magnified by 1000 and 3000 times respectively.

FIG. 2. X-ray diffraction patterns for the (a) random powders obtained by grinding the superconducting cores. The cross (+) denotes the observed diffraction intensity. The red and blue line denotes the fitted diffraction patterns and the difference between the observed and calculated intensity, respectively. Ag impurity phase marked with asterisks (*) was observed both in the rolled tapes and the random powders. (b) flat rolled $Ba_{0.6}K_{0.4}Fe_2As_2$ tape after peeling off the Ag sheath.

FIG. 3. (a) Temperature dependence of resistivity near the superconducting transition under a magnetic field (0, 1, 3, 5, 7 and 9 Tesla). Inset (c) shows the resistivity between 30 and 300 K. Inset (d) shows the superconducting transition width $\Delta T_c$, which is determined by $T_c^{onset} - T_c^{zero}$, as a function of the magnetic field. (b) Main panel shows the H-T phase diagram of the $Ba_{0.6}K_{0.4}Fe_2As_2$ tapes. The upper critical field $B_{c2}$ and the irreversibility field $B_{irr}$ were derived using the criterion of the 90% and 10% values of the normal resistivity $\rho(39 K)$. The black dashed line is a linear fitting. Inset (e) is the Arrhenius plots of $\rho$ at various magnetic fields. The black dashed lines are the fitting results using Eq.(4). Inset (f) shows $\ln\rho_0$ vs $U_0$ derived from the Arrhenius relation. The small fitting errors marked as the error bars guarantee the validity of the thermal activated flux flow model.

FIG. 4. Magnetic field dependence of the transport critical current density $J_c$ for the $Ba_{0.6}K_{0.4}Fe_2As_2$ tape. The inset depicts the temperature dependence of susceptibility under a 30 Oe magnetic field parallel to the tape surface.

FIG. 5. Optical image of the polished cross section of the $Ba_{0.6}K_{0.4}Fe_2As_2$ tapes magnified by (a) 50 and (b) 500 times. SEM images for the (c) cross section and (d) planar view of the tape. SEM images for the longitudinal section of the tape magnified by (e) 1300 and (f) 3000 times. The arrow in (e) indicate the tape axis.

Table 1: Rietveld Refinements for the Ba$_{0.6}$K$_{0.4}$Fe$_2$As$_2$ tapes.

| **Main phase**: BaK-122 | |
|---|---|
| Fraction | 98.4 % |
| Space group | I4/*mmm* |
| Fitted Formula | Ba$_{0.615}$K$_{0.385}$Fe$_2$As$_2$ |
| *a* (Å) | 3.9086(5) |
| *c* (Å) | 13.3088(1) |
| *V* (Å$^3$) | 203.320(5) |
| **Minor phase**: Ag | |
| Fraction | 1.6 % |
| Space group | Fm-3m |
| *a* (Å) | 4.0629(1) |
| **Reliability factor** | |
| *R$_p$* | 5.19 % |
| *R$_{wp}$* | 7.17 % |
| $\chi^2$ | 4.665 |

Fig. 1

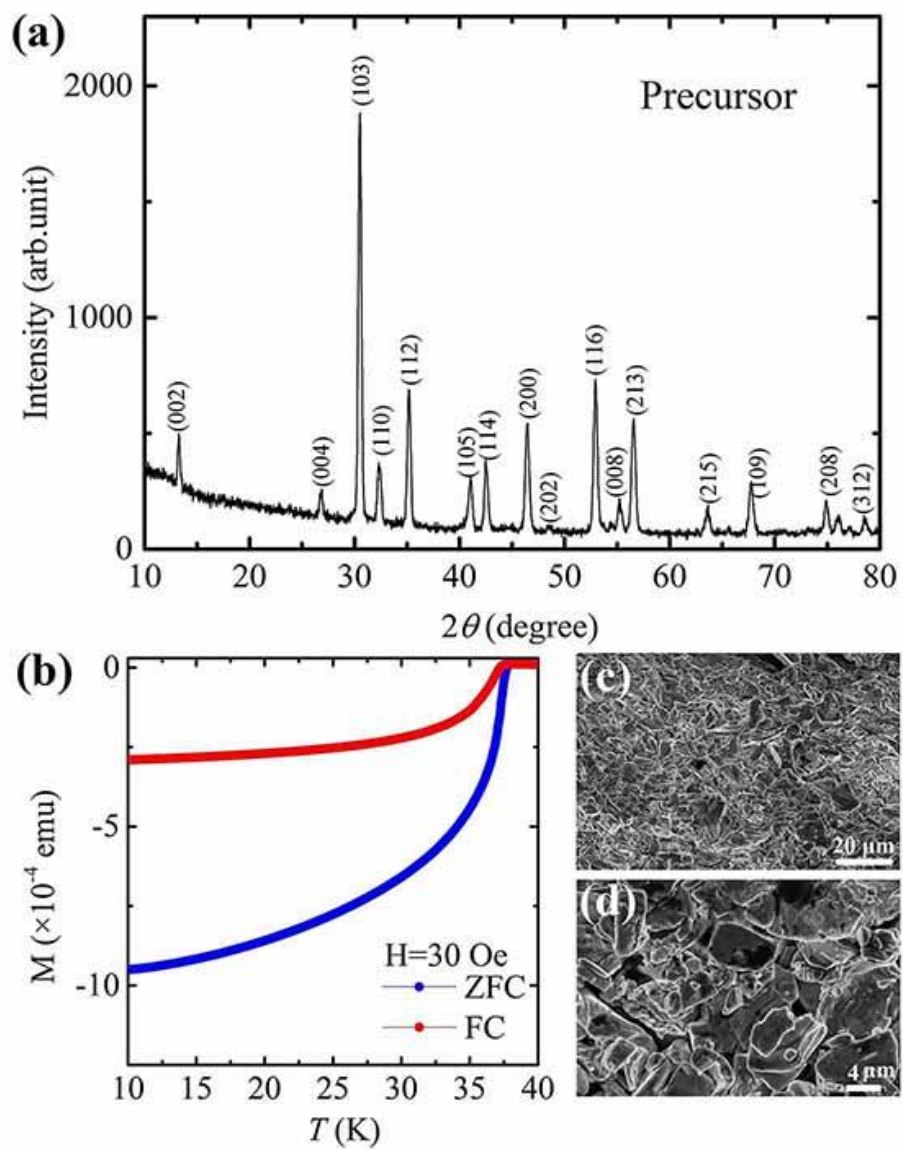

Fig. 2

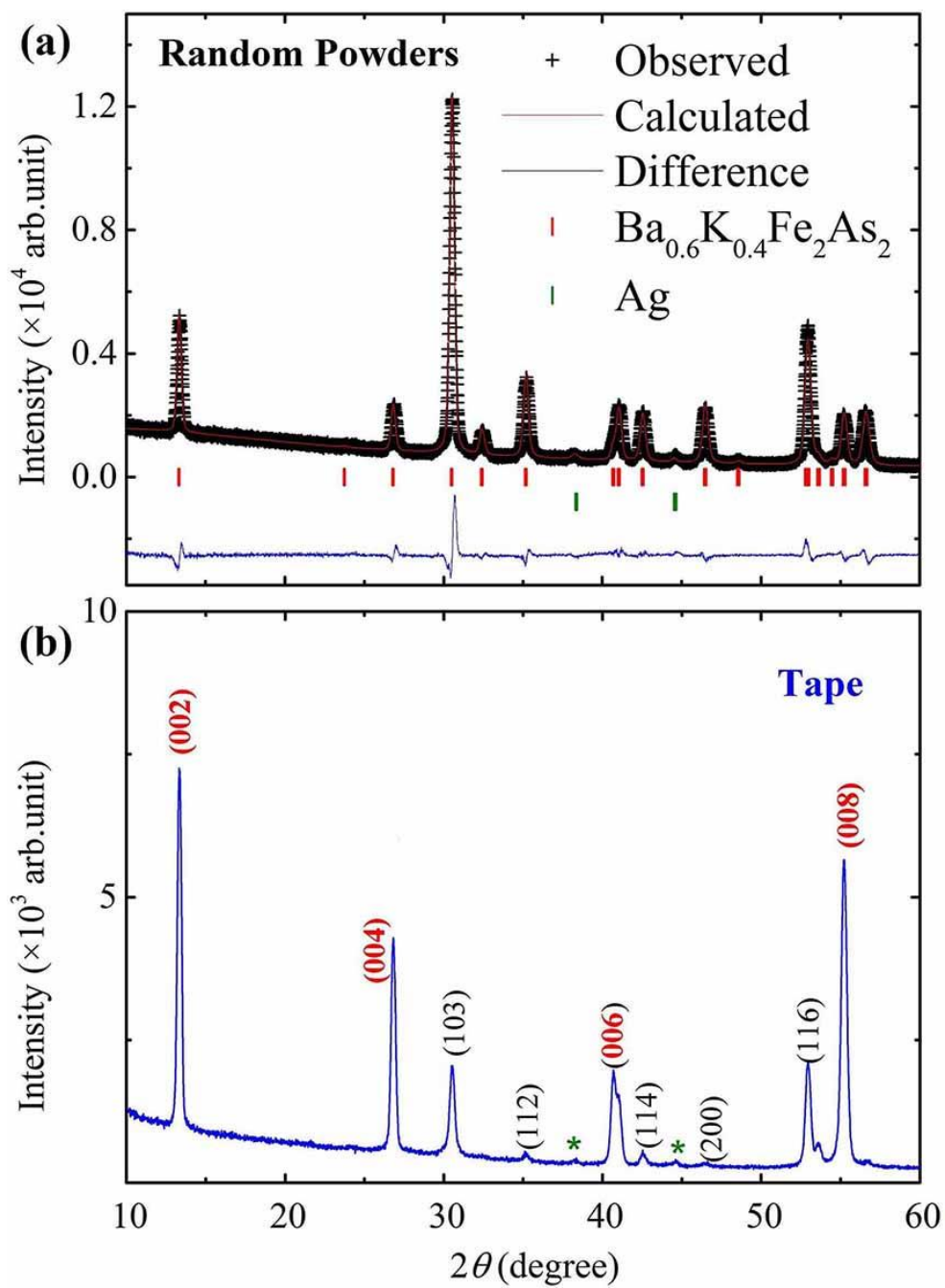

Fig. 3

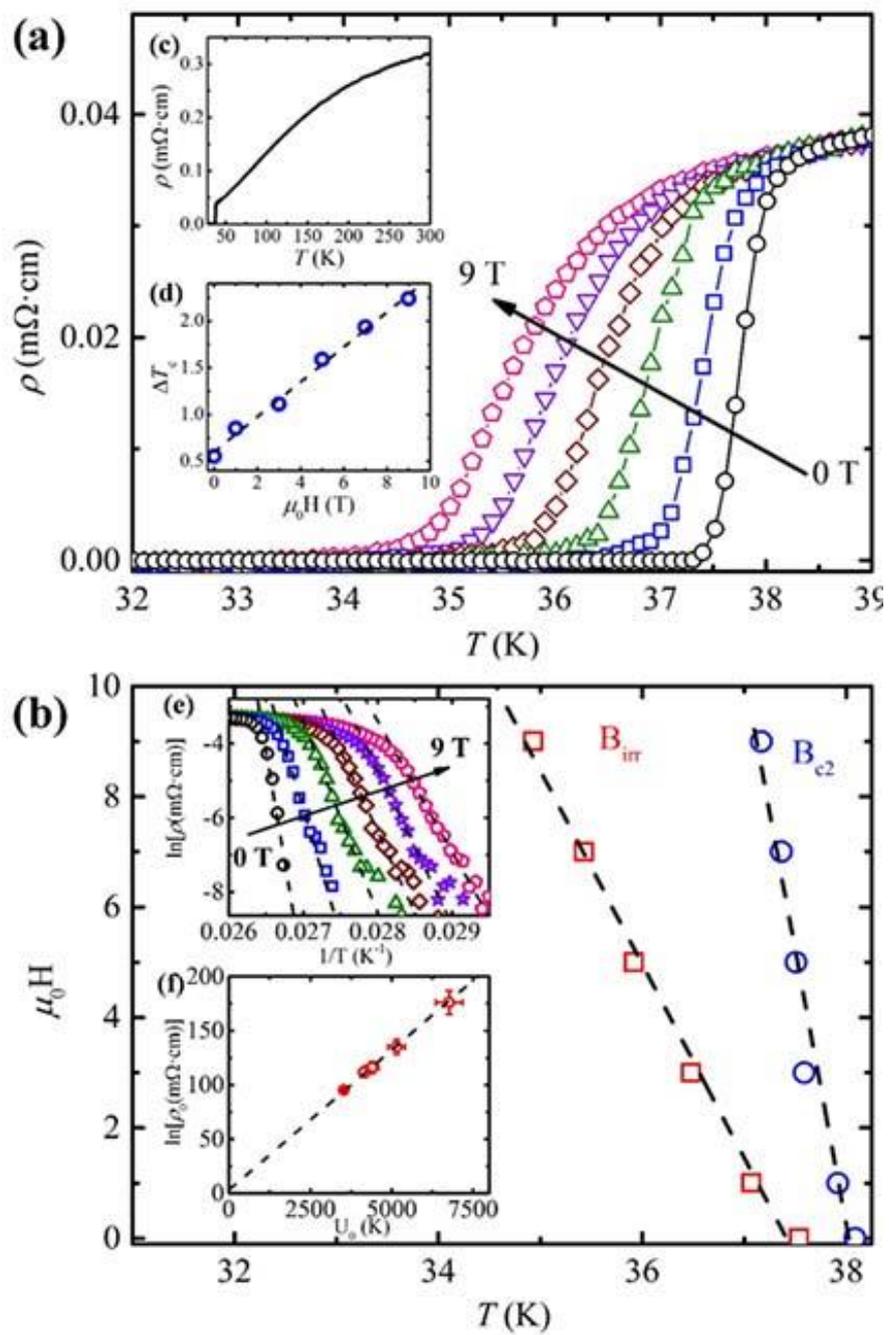

Fig. 4

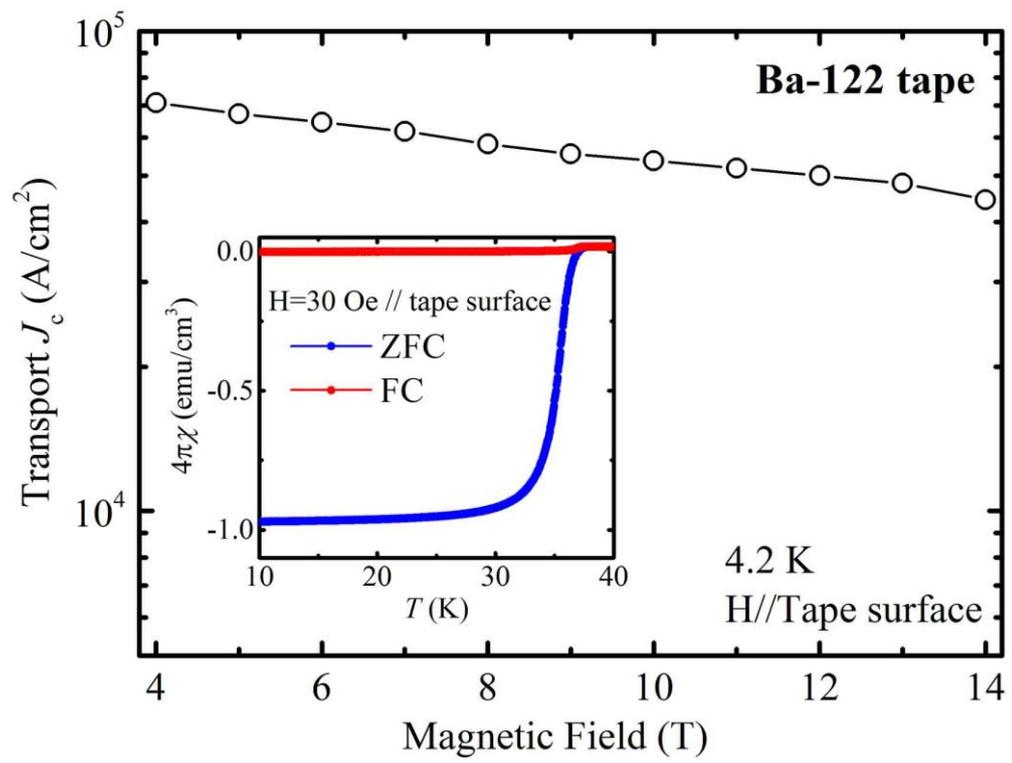

Fig. 5

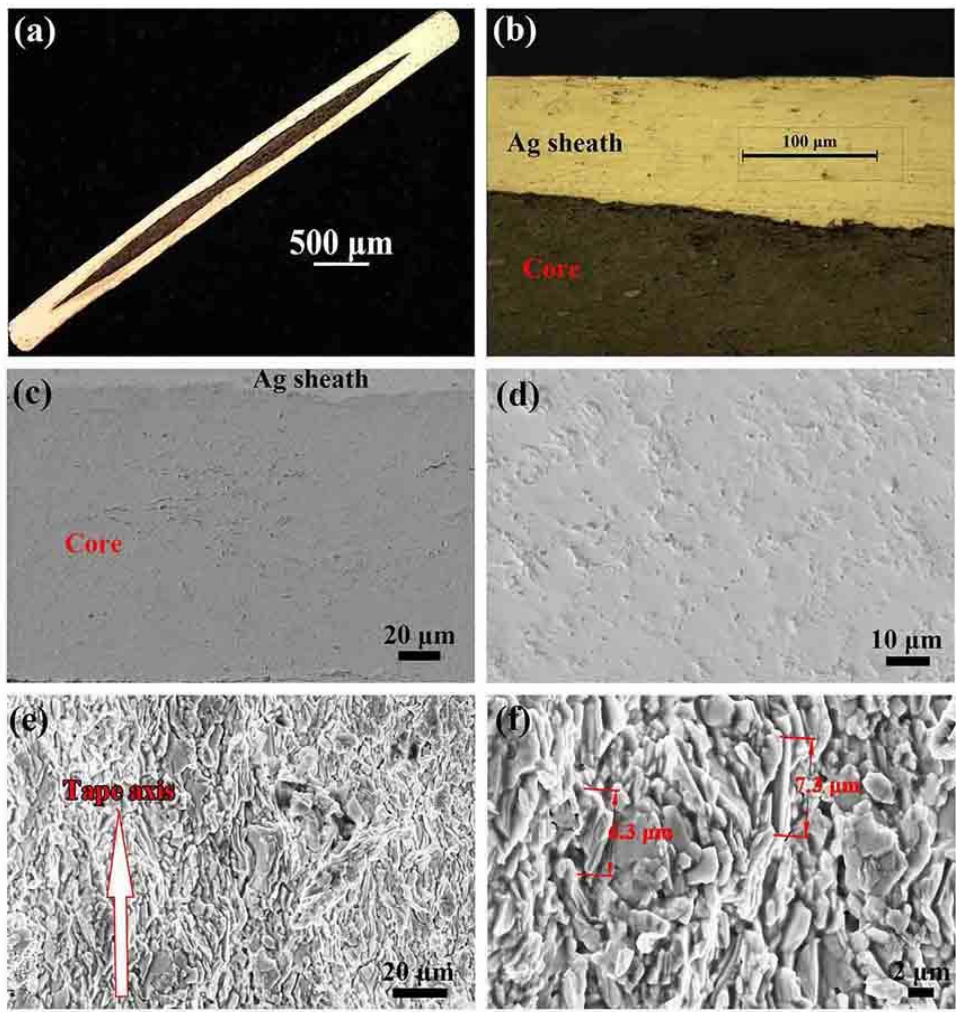